\documentclass{elsarticle}

\usepackage{lineno,hyperref,amsmath,nomencl,adjustbox,makecell,comment}
\usepackage[inline]{enumitem}
\modulolinenumbers[5]

\journal{Journal of \LaTeX\ Templates}

\usepackage{soul}
\usepackage[names]{xcolor}

\begin{document}

\begin{frontmatter}

\title{Comparison of behavioral systems theory and conventional linear models for predicting building zone temperature in long-term in situ measurements\tnoteref{mytitlenote}}

\author{Manuel Koch\textsuperscript{a,b} and Colin N. Jones\textsuperscript{a}}
\address{\textsuperscript{a}EPFL, 1015 Lausanne, Switzerland}
\address{\textsuperscript{b}FHNW, 4132 Muttenz, Switzerland}
\address{manuelpascal.koch@epfl.ch/manuel.koch@fhnw.ch, colin.jones@epfl.ch}




\begin{abstract}
The potential of Model Predictive Control in buildings has been shown many times, being successfully used to achieve various goals, such as minimizing energy consumption or maximizing thermal comfort. However, mass deployment has thus far failed, in part because of the high engineering cost of obtaining and maintaining a sufficiently accurate model. This can be addressed by using adaptive data-driven approaches. The idea of using behavioral systems theory for this purpose has recently found traction in the academic community. In this study, we compare variations thereof with different amounts of data used, different regularization weights, and different methods of data selection. Autoregressive models with exogenous inputs (ARX) are used as a well-established reference. All methods are evaluated by performing iterative system identification on two long-term data sets from real occupied buildings, neither of which include artificial excitation for the purpose of system identification. We find that: (1) Sufficient prediction accuracy is achieved with all methods. (2) The ARX models perform slightly better, while having the additional advantages of fewer tuning parameters and faster computation. (3) Adaptive and non-adaptive schemes perform similarly. (4) The regularization weights of the behavioral systems theory methods show the expected trade-off characteristic with an optimal middle value. (5) Using the most recent data yields better performance than selecting data with similar weather as the day to be predicted. (6) More data improves the model performance.

\end{abstract}


\end{frontmatter}

\newpage

\makenomenclature
\nomenclature{MPC}{Model Predictive Control}
\nomenclature{DeePC}{Data-enabled Predictive Control}
\nomenclature{BST}{Behavioral Systems Theory}
\nomenclature{ARX}{Autoregressive model with exogenous inputs}
\nomenclature{$\mathcal{B}$}{System}
\nomenclature{LTI}{Linear time-invariant}
\nomenclature{$n_{sys}$}{Order of system}
\nomenclature{$z$}{Input-output pair}
\nomenclature{$u$}{Control variable}
\nomenclature{$w$}{Disturbance}
\nomenclature{$y$}{Observation variable}
\nomenclature{$T$}{Number of samples}
\nomenclature{$T_{ini}$}{Initialization steps}
\nomenclature{$T_{f}$}{Prediction steps}
\nomenclature{$\mathcal{H}$}{Hankel/Trajectory matrix}
\nomenclature{PE}{Persistent excitation}
\nomenclature{HRV}{Heat recovery}
\nomenclature{$T_{z}$}{Zone temperature}
\nomenclature{PV}{Photovoltaics}
\nomenclature{DHW}{Domestic hot water}
\nomenclature{HP}{Heat pump}
\nomenclature{$P_{heat}$}{Heating power}
\nomenclature{$P_{cool}$}{Cooling power}
\nomenclature{$T_{amb}$}{Ambient temperature}
\nomenclature{$I_{sol}$}{Solar radiation}
\nomenclature{$\lambda$}{Regularization weight}
\nomenclature{$\alpha$}{Forgetting factor}
\nomenclature{$W_{I}$}{Initialization weighting matrix}
\nomenclature{$W_{H}$}{Width of trajectory matrix}
\nomenclature{RMSE}{Root mean square error}
\nomenclature{$t$}{Time}
\nomenclature{$k$}{Discrete time step}
\nomenclature{STD}{Standard deviation}
\nomenclature{Adpt.}{Adaptive}
\printnomenclature


\section{Introduction}

Buildings account for a quarter of total energy consumption globally and up to $40 \%$ in developed countries; the majority of which is used for heating (including hot water), cooling and ventilation. Electric heating with heat pumps is slowly displacing the use of oil and natural gas \cite{perez-lombardReviewBuildingsEnergy2008, urge-vorsatzHeatingCoolingEnergy2015}. Meanwhile, the total energy consumption for cooling is expected to grow substantially, especially in developing countries \cite{isaacModelingGlobalResidential2009, FutureCoolingAnalysis}. 

Model Predictive Control (MPC) has been successfully applied to buildings in simulations and experiments many times \cite{seraleModelPredictiveControl2018,aframTheoryApplicationsHVAC2014,mirakhorliOccupancyBehaviorBased2016,drgonaAllYouNeed2020}. While the most common objectives in these studies are to reduce the energy consumption or to increase thermal comfort, there is an increasing focus on harnessing buildings for electric demand response. Considering the ongoing electrification of heating and the increase in cooling on one side, as well as the increasing share of non-dispatchable sources of electricity like wind and solar on the other side, this trend can be expected to continue. Unlike the first two objectives, which can be achieved through predictive and non-predictive methods, demand response usually requires a reasonably accurate prediction of the electricity demand several hours into the future \cite{qureshiModelPredictiveControl2014}. 

However, MPC is yet to find widespread commercial application in buildings. A commonly cited reason is the high upfront engineering cost of identifying a sufficiently accurate model of the building dynamics \cite{seraleModelPredictiveControl2018,killianTenQuestionsConcerning2016,woliszSelflearningModelPredictive2020,sturzeneggerModelPredictiveClimate2016}. 

Buildings have a number of properties that inform the selection of an appropriate model structure: Unlike most industrial products, they are generally unique, which means a different model must be built for each building. There is also a great variety in size and complexity, ranging from a single thermal zone to hundreds. While we usually measure and control the air temperature in the occupied zones, which have time constants in the range of minutes, most of the thermal energy is stored in the walls, floors and furniture, which can have time constants in the range of hours. 
Furthermore, buildings show slowly time-varying behavior, including cyclical changes between seasons, but also long-term changes due to degrading insulation or altered furnishing. However, the instantaneous dynamics of a building can be modeled as an LTI system with sufficient accuracy \cite{seraleModelPredictiveControl2018, drgonaAllYouNeed2020}. 
The execution of system identification experiments with high excitation of the heating and cooling systems is constrained by a desire to maintain occupant comfort during occupied hours and to minimize strain on the HVAC components. 
Despite these challenges, we desire to obtain reasonably accurate predictions over a $24 h$ horizon, with the goal to participate in demand response. 
The intention to eventually deploy our work at scale outside of academic settings adds additional challenges: Sensors in buildings are often inaccurate and biased, and a replacement or calibration may not be an option. In addition, excessive computational requirements should be avoided to keep the system hardware costs as low as possible. 

To address these issues, we propose the use of adaptive, linear black-box models \cite{maddalenaDatadrivenMethodsBuilding2020,hilliardModelPredictiveControl2016}. In particular, we investigate the performance of a recently popularized approach based on behavioral systems theory, further detailed in sections \ref{SecDeePC} and \ref{DeePCdetail}. These approaches omit the step of model identification altogether. However, they have a theoretical equivalence to linear time-invariant models. Therefore, we choose autoregressive models with exogenous inputs (ARX) as a well-established reference method. Furthermore, we evaluate non-adaptive variants of both methods to assess the benefits of adaptivity. 

Other methods have been used to model buildings in the existing literature, but were disregarded for the scope of this paper. Neural networks and Gaussian process models can account for nonlinearities, but they require large amounts of data to train and a computationally expensive nonlinear optimization problem to be solved. This contradicts our goal to identify a mostly linear system with limited computational effort. A combination of a state-space model and a Kalman filter for state estimation was considered as an alternative to ARX, but not chosen due to the higher complexity with no obvious advantages in this context.

This study evaluates the ability of the selected modeling approaches to predict the zone temperature with sufficient accuracy on two long-term measurements from real occupied residential buildings, detailed in section \ref{SecData}. The data stems from regular operation, meaning that there is no active excitation for the purpose of system identification. Furthermore, the data contains gaps in the measurements and unknown levels of process and measurement noise. Testing under these challenging conditions is essential to draw conclusions for a future deployment in real-world buildings at scale. 

\section{Theoretical background}
\subsection{Data-enabled predictive control}\label{SecDeePC}
Data-enabled predictive control (DeePC) is a formulation of MPC based on behavioral systems theory. It was first published and named by Coulson in 2019 \cite{coulsonDataEnabledPredictiveControl2019}. Unlike classical MPC, no model of the plant is identified. The solution to the predictive control problem is generated directly by a linear combination of measured trajectories of the plant. This section briefly explains the method: 

Assume a linear, time-invariant system $\mathcal{B}$ of order $n_{sys}$ with discrete-time input-output samples $z_k = $
$\begin{bmatrix}
u_k & y_k
\end{bmatrix}
^\top$
and a measurement of $T$ samples $z_{[1:T]}$. Willems' fundamental lemma \cite{willemsNotePersistencyExcitation2005} shows that any possible trajectory of system $\mathcal{B}$ of length $T_f$ can be constructed as a linear combination of the columns of the Hankel matrix of depth $L=n_{sys} + T_f$:

$\mathcal{H}_L(z_{[1:T]}) = \begin{bmatrix}
z_1 & \dots & z_{T-L+1}\\
\vdots & \ddots & \vdots\\
z_L & \dots & z_T
\end{bmatrix}$
if $\mathcal{H}_L(u_{[1:T]})$ has full row rank. 

Now consider the Hankel matrices $\mathcal{H}_L(u_{[1:T]}) = \begin{bmatrix}
\mathcal{H}_{T_{ini}}^u\\
\mathcal{H}_{T_{f}}^u
\end{bmatrix}$
and $\mathcal{H}_L(y_{[1:T]}) = \begin{bmatrix}
\mathcal{H}_{T_{ini}}^y\\
\mathcal{H}_{T_{f}}^y
\end{bmatrix}$
each split into two components for initialization and prediction and combined to $\mathcal{H}_L(z_{[1:T]}) = \begin{bmatrix}
\mathcal{H}_{T_{ini}}^u\\
\mathcal{H}_{T_{ini}}^y\\
\mathcal{H}_{T_{f}}^u\\
\mathcal{H}_{T_{f}}^y
\end{bmatrix}$.

As well as the right-hand vector $v = \begin{pmatrix} u_{ini} \\ y_{ini} \\ u \\ y \end{pmatrix}$, consisting of the $T_{ini}$ most recent samples $u_{ini}$ and $y_{ini}$, for initialization and the optimization variables $u$ and $y$. 

With $\mathcal{H}_L(z_{[1:T]})$ simplified to $\mathcal{H}$ and the introduction of the optimization variable $g$, which represents the linear combination of the columns of $\mathcal{H}$, the predictive controller is formulated: 

\begin{align*}
&\min_{g, u, y} J(u, y) \\
&\text{subject to} \nonumber \\
&\mathcal{H} g = v \\
&u \in \mathcal{U} , y \in \mathcal{Y}
\end{align*}

Many studies employing this method have been published since it was first proposed. A recent literature review is given in \cite{markovskyBehavioralSystemsTheory2021a}. 

\subsection{Persistency of excitation for different trajectory matrices}
Persistency of excitation (PE) is a condition on a measured data set for the underlying system to be identifiable. Willems' fundamental lemma formulates this condition for data in Hankel matrix form. Subsequent publications have extended this to mosaic-Hankel matrices and Page matrices, shown in Tab. \ref{tab:PEconditions}. A mosaic-Hankel matrix is a horizontal concatenation of multiple Hankel matrices. A Page matrix is similar to a Hankel matrix, but does not have repeat entries. It instead contains a continuous string of samples, similar to words on a page, hence the name. We note that the very strict condition for the Page matrix is sufficient but not necessary. In fact, a quadcopter is successfully controlled with a Page matrix in \cite{coulsonDistributionallyRobustChance2021}, despite grossly violating the stated PE condition derived in that paper. This serves as an example of the importance of experimental studies complementing the theoretical work recently published in this field. 

In our study, we use Hankel-like matrices comprised of individual, disjointed, but overlapping trajectories. To the best of our knowledge, no PE condition has been formalized for such an unstructured matrix. Considering the aforementioned findings, we conduct a study on real data without an adapted PE condition available. 

\begin{table}[]
\begin{tabular}{l | l | l}
Matrix        & PE condition                                                             & Minimal $T$                                                                                  \\ \hline
Hankel \cite{coulsonDistributionallyRobustChance2021}       & $\mathcal{H}_L(u_{[1,T]})$ has full row rank                                                      & \begin{tabular}[c]{@{}l@{}}$T \geq L(m+1)-1$\\ is of order $L$\end{tabular}                            \\ \hline
mosaic-Hankel \cite{vanwaardeWillemsFundamentalLemma2020}& \begin{tabular}[c]{@{}l@{}}$\begin{bmatrix}
    \mathcal{H}_k(u_{[0,T_1-1]}^1) & \mathcal{H}_k(u_{[0,T_2-1]}^2) & \dots & \mathcal{H}_k(u_{[0,T_q-1]}^q)
    \end{bmatrix}$\\ has full row rank\end{tabular}  & \begin{tabular}[c]{@{}l@{}}$\sum\limits_{i=1}^q T_i \geq k(m+q)-q$\\ is of order k\\ $len(T_i) \geq k$\\ Necessary condition\end{tabular} \\ \hline
Page \cite{coulsonDistributionallyRobustChance2021}         & \begin{tabular}[c]{@{}l@{}}$\begin{pmatrix}
    \mathcal{P}_L(u_{[1, T-(M-1)L]}) \\
    \mathcal{P}_L(u_{[L+1, T-(M-2)L]}) \\
    \vdots \\
    \mathcal{P}_L(u_{[L(M-1)+1, T]}) \\
    \end{pmatrix}$\\ has full row rank\end{tabular} & \begin{tabular}[c]{@{}l@{}}$T \geq L((mL+1)M-1)$\\ is of order k\\ Sufficient condition\end{tabular}    
\end{tabular}
\caption{Persistent excitation condition and minimal number of sampling points for different trajectory matrix structures}
\label{tab:PEconditions}
\end{table}

\subsection{Persistent excitation in dual control} \label{PeDualControl}
In case of an adaptive scheme, where the Hankel matrix (or any other data structure) is continuously updated, continuously ensuring persistent excitation is a challenge. The published methods to increase the level of excitation for adaptive data-driven controllers can be grouped into three categories: The first is selecting suitable data from a long data log. The second is adding a perturbation to the solved predictive control trajectory. The third is including the level of excitation in the cost function \cite{berberichLinearTrackingMPC2021}. 
However, buildings are significantly perturbed by the weather, which cannot be controlled. This favors the first method. Furthermore, artificially increasing the level of excitation of the control signal may increase wear on the actuators as well as constraint violations. In the case of buildings, the resulting temperature fluctuations may also decrease occupant comfort. 

\subsection{DeePC in building control}
DeePC is derived for deterministic, linear, time-invariant systems. Since none of these conditions perfectly apply to buildings, experiments and high-fidelity simulations are necessary to evaluate the practically achievable level of performance. 

The existing literature is sparse. Schwarz \cite{schwarzDataDrivenControlBuildings2020}, O’Dwyer \cite{odwyerDatadrivenPredictiveControl2021} and Chinde \cite{chindeDataEnabledPredictiveControl} have conducted simulation studies on DeePC in buildings. However, these are not adaptive. Kerkhof \cite{kerkhofOptimalControlAutonomous2020} has conducted a simulation study on adaptive DeePC in a greenhouse. However, a very short prediction horizon of one hour is used. Also, the dynamics of a greenhouse may be very different from those of a conventional building. Lian \cite{lianAdaptiveRobustDatadriven2021} has conducted an experimental study on adaptive robust bi-level DeePC in an educational building. Kerkhof and Lian both add artificial noise to the control signal to ensure persistent excitation.

\section{Methodology}
Since DeePC refers to a control method and we focus on system identification, the methods based on behavioral systems theory are referenced with the label ''BST'' hereinafter, rather than ''DeePC''. For each building, a total of 69 modeling methods are evaluated: One non-adaptive ARX, three adaptive ARX with different forgetting factors, five non-adaptive BST with different regularization weights, and 60 adaptive BST with different trajectory matrix widths, regularization weights and data selection methods. Non-adaptive variants are identified before the evaluation phase. Adaptive variants are updated at every step, starting with a 30 day initialization phase before the evaluation phase. For all methods, 12 steps are used for initialization and 96 steps for prediction, corresponding to $3 h$ and $24 h$. 

The weather forecasts for the prediction are ideal. However, a manipulated forecast is used for the data selection for the adaptive BST variants, detailed in section \ref{WeaManMet}. 

As is typical for long-term field measurements in buildings, there are many gaps in the data where sensors returned no measurement or physically impossible values. They are identified in the pre-processing. Uninterrupted trajectories of length $T_{ini}+T_f$ or greater are marked as admissible for system identification and validation. Inadmissible segments are skipped. For the adaptive ARX, the model update is halted when encountering a gap. For the non-adaptive ARX, a multi-batch identification method is used in case of gaps. 

\subsection{BST variants in detail}\label{DeePCdetail}

A prediction of the zone temperature, based on the current state and a weather forecast, is performed at every time step. The control variables are the heating and cooling powers $u = \begin{bmatrix} P_{heat} & P_{cool} \end{bmatrix}$ (heating only for Basel building). The disturbance variables are the ambient temperature and solar radiation $w = \begin{bmatrix} T_{amb} & I_{sol} \end{bmatrix}$. The output variable is the zone temperature $y = T_z$. With the disturbance variable, the DeePC formulation from section \ref{SecDeePC} changes to: 

$\mathcal{H}_L(z_{[1:T]}) = \begin{bmatrix}
\mathcal{H}_{T_{ini}}^u\\
\mathcal{H}_{T_{ini}}^w\\
\mathcal{H}_{T_{ini}}^y\\
\mathcal{H}_{T_{f}}^u\\
\mathcal{H}_{T_{f}}^w\\
\mathcal{H}_{T_{f}}^y
\end{bmatrix}$ and $v = \begin{pmatrix} u_{ini} \\ w_{ini} \\ y_{ini} \\ u \\ w_{forecast} \\ y \end{pmatrix}$

We further define $\hat{\mathcal{H}} = \begin{bmatrix}
\mathcal{H}_{T_{ini}}^u\\
\mathcal{H}_{T_{ini}}^w\\
\mathcal{H}_{T_{ini}}^y\\
\mathcal{H}_{T_{f}}^u\\
\mathcal{H}_{T_{f}}^w
\end{bmatrix}$ the first five elements of $\mathcal{H}_L(z_{[1:T]})$ and $\hat{v} = \begin{pmatrix} u_{ini} \\ w_{ini} \\ y_{ini} \\ u \\ w_{forecast} \end{pmatrix}$ the first five elements of $v$.

The following cost function comprised of a least-squares fit and a regularization on $g$ is minimized: 

\begin{equation}
    J = (\hat{\mathcal{H}} g - \hat{v})^\top W_{I} (\hat{\mathcal{H}} g - \hat{v}) + \lambda g^\top g
    \label{eq:CostFunction}
\end{equation}

Which is solved analytically to: 

\begin{equation}
    g^* = (\hat{\mathcal{H}}^\top W_{I} \hat{\mathcal{H}} + \lambda I)^{-1} \hat{\mathcal{H}}^\top W_{I} \hat{v}
\end{equation}

\begin{equation}
    y^* = \mathcal{H}_{T_{f}}^y g^*
\end{equation}

$W_{I}$ is a weighting matrix that gives the initialization steps 100 times the weight of the prediction steps. This is done to improve the accuracy of the early prediction steps, which are more critical for predictive control, since usually only the first element of the optimized trajectory is applied, before the entire optimization is repeated. 

The widths of the trajectory matrices, denoted $W_H$, for the adaptive variants are $181$, $373$ and $661$. In a proper Hankel matrix, this would correspond to $3$, $5$ and $8$ days of data $(W_H = T - T_{ini} - T_f + 1)$. For the non-adaptive BST, all data from the identification phase is used to build one matrix several 1000 columns wide.

The regularization weights $\lambda$ are $10^0$, $10^1$, $10^2$, $10^3$ and $10^4$. 

\subsubsection{Data selection methods in detail}\label{DatSelDetail}

\begin{figure}[htp]
    \centering
    \includegraphics[width=8cm]{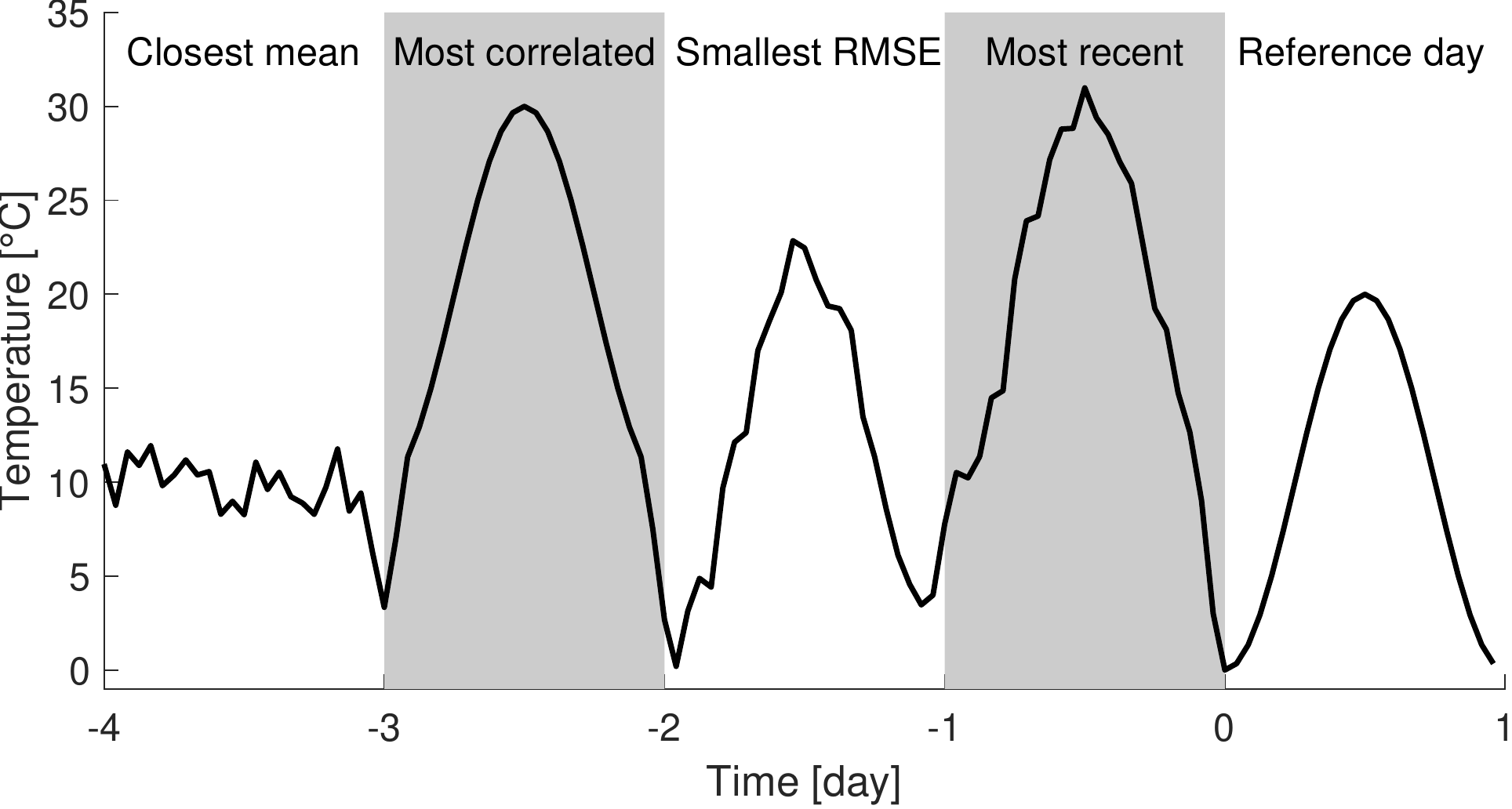}
    \caption{Simple visualization of the data selection variants. Four past days are compared to one forecast reference day and categorized into closest mean, most correlated, smallest RMSE and most recent.}
    \label{fig:DatSelVar}
\end{figure}

We investigate the impact of different ways of composing the trajectory matrix. One method that is as close as possible to a proper Hankel matrix and three alternative approaches. Following, the four different methods used are explained. A simplified visualization is given in Fig. \ref{fig:DatSelVar}.

\textbf{\textit{Most recent:}} The $W_H$ most recent admissible trajectories are selected. If there are no gaps in the data, this yields a proper Hankel matrix. If there are gaps, we get a mosaic-Hankel matrix. 

\textbf{\textit{Most correlated:}} The $W_H$ admissible trajectories whose weather is most correlated with the forecast are selected. This method serves the purpose of investigating if a selection of trajectories with a shape similar to the target trajectory is advantageous, compared to the aforementioned approach. 

\textbf{\textit{Smallest RMSE:}} The $W_H$ admissible trajectories whose weather has the smallest RMSE relative to the forecast are selected. This method is similar to the correlation one, but is not invariant to scaling and offset. It is similar to a parameter-variant model, with the weather as the parameter in question. Due to the multi-step nature of the BST approach, a possible parameter variance would be captured implicitly, rather than explicitly modeled. Since ambient temperature and solar radiation generally follow a daily pattern, it also includes an element of time-variant modeling. 

\textbf{\textit{Closest mean:}} The $W_H$ admissible trajectories whose averaged weather most closely match the corresponding values of the forecast are selected. This method seeks to strike a balance between selecting recent data and selecting similar data. Because average ambient conditions typically change on the scale of weeks and months, it is expected to be similar to selecting the most recent data. 

Data selection is limited to the past 365 days for all methods. Since the calculation is initialized with roughly 30 days of data, the amount of available data increases with time, until the one year window is full. While a constant amount of data over the calculation period would be preferable, it is not possible due to the limited data sets. 

The weather data consists of the normalized ambient temperature and solar radiation. The sum of the two correlation coefficients, RMSE and differences is used. The ambient temperature is normalized to $\hat{T}_{amb} = \frac{T_{amb} - 10 [K]}{20 [K]}$. This compresses the range $[- 10 ^{\circ} C, 30 ^{\circ} C]$ to $[-1, 1]$. For the solar radiation, the normalization is adapted to the specific sensor situations detailed in section \ref{SecData}. $\hat{I}_{sol} = I_{sol}/(500 [W/m^2])$ for Zürich and $\hat{P}_{PV} = P_{PV}/(3 [kW])$ for Basel. These values are chosen to rescale the values to roughly $[0, 1]$. The different normalization ranges counteract the fact that the ambient temperature usually stays within a narrow segment of the normalization range during $24 h$. The solar radiation, on the other hand, regularly spans much of its normalization range during one summer day. 

Despite these methods not yielding actual Hankel matrices, we denote them all with the symbol $\mathcal{H}$ for simplicity.

\subsubsection{Manipulated weather forecast for data selection}\label{WeaManMet}
There is a key difference between the variants based on selecting similar weather and the others: They are impacted by the inaccuracy of the weather forecast twice. Once in the selection step and once in the prediction step. While we use an ideal forecast for all variants in the prediction, which creates an equal comparison, we manipulate the forecast for the data selection step to simulate the impact of this double-dependency. 

To the best of our knowledge, there is no readily available tool to generate realistic artificial forecasts based on measured data. Therefore, we defined our own method. While a real (or at least realistic) forecast would have been preferred, we consider this to reasonably serve the purpose of simulating the inaccuracy of a real forecast for the scope of this study. For the same reason, \cite{scharnhorstEnergymBuildingModel2021} relies on a self-made method for skewing of the weather data as well. 

\begin{figure}[htp]
    \centering
    \includegraphics[width=8cm]{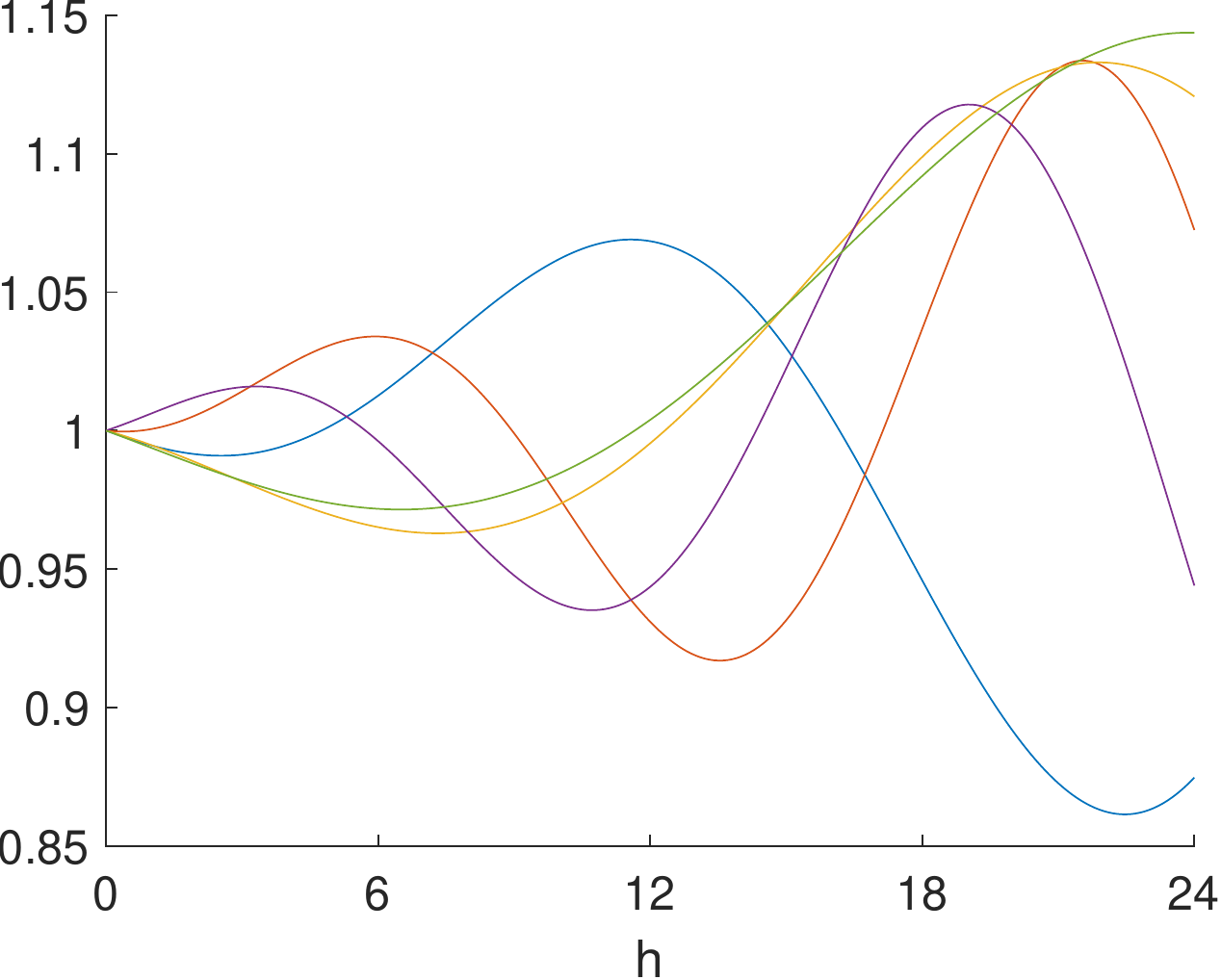}
    \caption{Five example trajectories of the distortion of the weather forecast}
    \label{fig:distortWeather}
\end{figure}

A randomized sine function is generated, that grows linearly over the prediction horizon of $24 h$. For the ambient temperature, it is used in an additive way and grows from $0 K$ to a maximum of $\pm 4 K$. For the solar radiation, it is used in a multiplicative way and grows from $1$ to a maximum $\pm 15 \%$. Fig. \ref{fig:distortWeather} shows some example trajectories for the multiplicative case. 

\begin{align*}
& f_{scale} = 1 + 0.15 \frac{t}{24} sin(a t + b) \\
& \text{with} \nonumber \\
& a = \mathcal{U}(0.5 \frac{\pi}{12}, 1.5 \frac{\pi}{12}) \text{(Uniform distribution)} \\
& b = \mathcal{U}(-\pi, \pi) \\
& t = (0, 24) \text{(Hour of day)}
\end{align*}

\subsection{ARX variants in detail}

An ARX model has the following structure \cite{EstimateParametersARX}: 

\begin{equation}
    y(k) = a_1 y(k-1) + ... + a_{n_a} y(k - n_a) = b_1 u(k - n_k) + ... + b_{n_b} u(k - n_b - n_k + 1)
\end{equation}

With $n_a$ the number of poles, $n_b$ the number of zeros and $n_k$ the number of input samples that occur before the input affects the output. The parameters are estimated by a least-squares fit. The prediction over the whole horizon is iterative by chaining the one-step prediction model $T_f$ times. 

For the adaptive ARX, a forgetting factor method is used \cite{MathworksRecursiveArx}: 

\begin{equation}
    \hat{\Theta}(k) = \hat{\Theta}(k - 1) + K(k)(y(k) - \hat{y}(k))
\end{equation}

Where $\hat{\Theta}(k)$ is the parameter estimate at time k, $y(k)$ is the measurement and $\hat{y}(k)$ is the prediction thereof. The gain has the form: 

\begin{equation}
    K(k) = Q(k) \Psi(k)
\end{equation}

Where $\Psi(k)$ is the gradient of the predicted output $\hat{y}(k)$ with respect to $\Theta$. We further define: 

\begin{equation}
    Q(k) = \frac{P(k - 1)}{\alpha + \Psi^\top(k) P(k - 1) \Psi(k)}
\end{equation}

and

\begin{equation}
    P(k) = \frac{1}{\alpha} \left( P(k - 1) - \frac{P(k - 1) \Psi(k) \Psi(k)^\top P(k - 1)}{\alpha + \Psi(k)^\top P(k - 1) \Psi(k)} \right)
\end{equation}

\begin{equation}
    P(0) = 10000
\end{equation}

With the forgetting factor $\alpha$. The specific values of $\alpha$ used are chosen to have time constants of $3$, $5$ and $8$ days, based on the relationship $\alpha = 1 - 1/T$. $P(0)$ is the default value of the used Matlab function. 

\subsection{Analysis}
The prediction quality is quantified by the root mean square error (RMSE) between the prediction and the actual values. For the error trajectories, the mean and standard deviation (STD) of the absolute error at each prediction step is used. 

\section{Data sets} \label{SecData}
Two data sets are used. The first one is an apartment near Zürich, completed in 2018 \cite{dinatalePhysicallyConsistentNeural2021}. It is located on the second floor of a small vertically stacked research neighborhood. Its construction emphasizes the use of wood and recycled materials, as well as experimental construction materials derived from fungi. It consists of a living room, two bedrooms, two bathrooms and a small entrance space. For the study, only the living room and the bedrooms, which have a total area of $93.8 m^2$, are considered. The temperatures in these rooms are measured and averaged. Heating and cooling are provided by a radiant slab with a central supply of hot and cold water. The thermal powers are measured at the connection points to the apartment. The ambient temperature and the solar radiation are measured locally. The pyranometers to measure the solar radiation are aligned with the orientation of the windows. The apartment has one large window front with a north-east orientation, sandwiched between overhanging concrete slabs. Therefore, the solar gains strongly depend on the position of the sun and can differ substantially from the global horizontal radiation. 

\begin{figure}[htp]
    \centering
    \includegraphics[width=8cm]{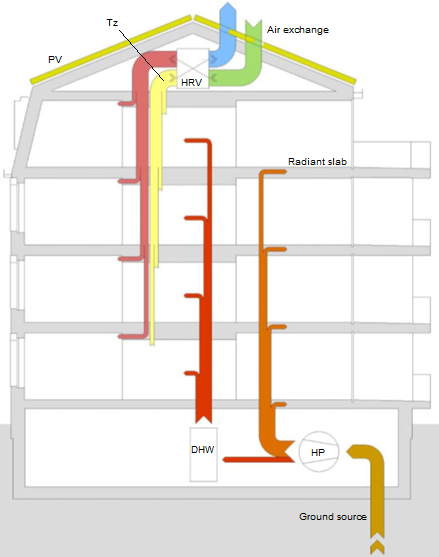}
    \caption{Basel building plan}
    \label{fig:BaselPlan}
\end{figure}

The second location is a terraced apartment building in Basel, completed in 2015 \cite{MonitoringMinergieAEcoMFH}. It consists of seven units with 2.5 to 3.5 rooms. It is a low-energy building with a measured thermal heating energy demand of $17 kWh/(m^2a)$. It is heated by radiant slabs, supplied by a ground-source heat pump. Since only the total electric power consumption of the heat pump is measured, a lumped internal temperature of the whole building, measured in the collective ventilation exhaust before the heat recovery unit is used, marked as $T_z$ in Fig. \ref{fig:BaselPlan}. The ambient temperature is locally measured. Because there is no local measurement of the solar radiation, the power of the roof-top photovoltaic installation with a north-west/south-east orientation is used as a substitute. Due to its size, split orientation and mostly unobstructed location, it is assumed to be a good approximation of the global horizontal radiation. The windows are moderately sized and well-shaded to minimize overheating in the summer. 

Both data sets span one year and are sampled with a $15$ min time step. The raw data is measured at a higher frequency and resampled. This allows to eliminate sensor noise in the pre-processing. This method could also be used in real-time. 

Roughly $365$ days of data are selected for the evaluation, plus roughly $30$ days of initialization for the adaptive variants and $30$ days each of heating (and cooling) operation for the system identification of the non-adaptive variants. Depending on the presence of heating and cooling, and to compensate for gaps in the data, these numbers are slightly adjusted, shown in Tab. \ref{tab:ExpDat}. The missing data refers to the evaluation period. We note that one missing data point prevents multiple time steps from being evaluated, based on the lengths of the initialization and horizon of the prediction. 

\begin{table}[]
\begin{tabular}{l|c|c}
                       & Zürich building         & Basel building          \\ \hline
Heating identification & 17.02.2020 - 17.03.2020 & 02.10.2016 - 22.11.2016 \\ \hline
Cooling identification & 21.08.2020 - 19.09.2020 & 
no cooling                        \\ \hline
Initialization         & 21.08.2020 - 20.09.2020 & 02.10.2016 - 22.11.2016 \\ \hline
Evaluation                   & 21.09.2020 - 21.09.2021 & 23.11.2016 - 23.11.2017 \\ \hline
Missing data                   & $\sim 10 \%$ & $\sim 2.5 \%$
\end{tabular}
\caption{Data selection and gaps}
\label{tab:ExpDat}
\end{table}

\begin{figure}[htp]
    \centering
    \includegraphics[width=8cm]{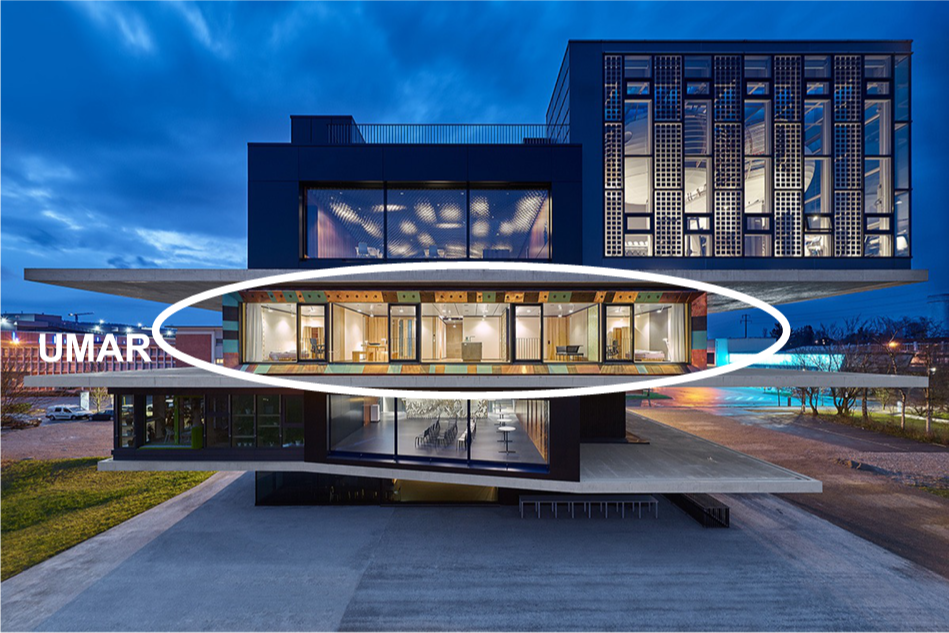}
    \caption{Zürich building}
    \label{fig:ZurichPhoto}
\end{figure}

\begin{figure}[htp]
    \centering
    \includegraphics[width=8cm]{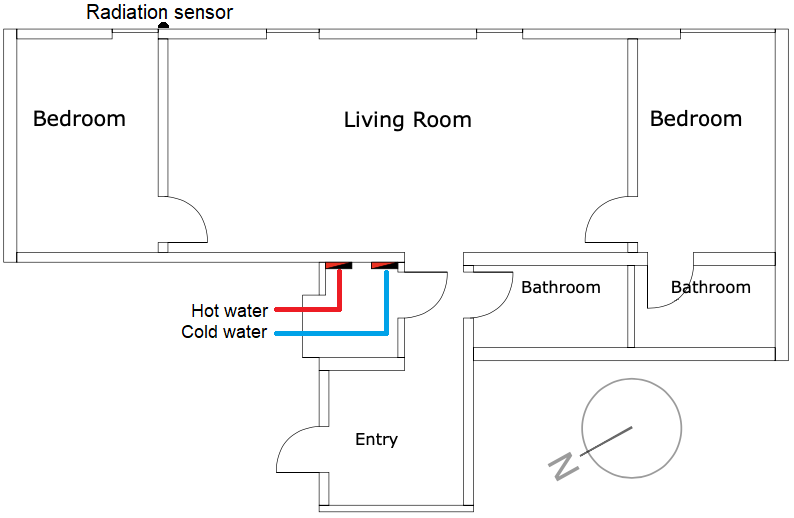}
    \caption{Zürich building plan}
    \label{fig:ZurichPlan}
\end{figure}

\begin{figure}[htp]
    \centering
    \includegraphics[width=8cm]{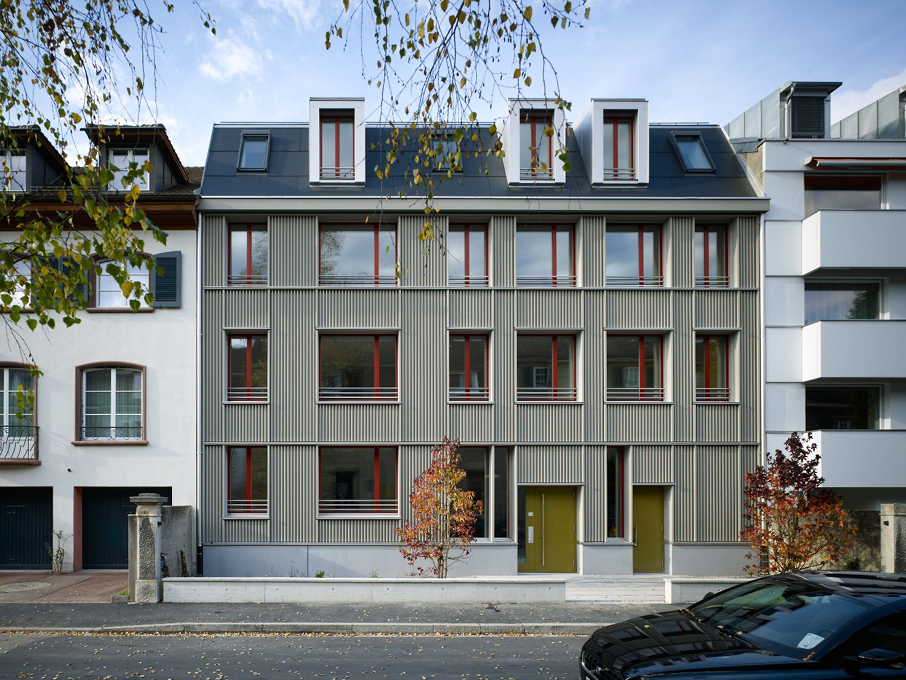}
    \caption{Basel building}
    \label{fig:BaselPhoto}
\end{figure}

\section{Results}

\begin{figure}[htp]
    \centering
    \includegraphics[width=12cm]{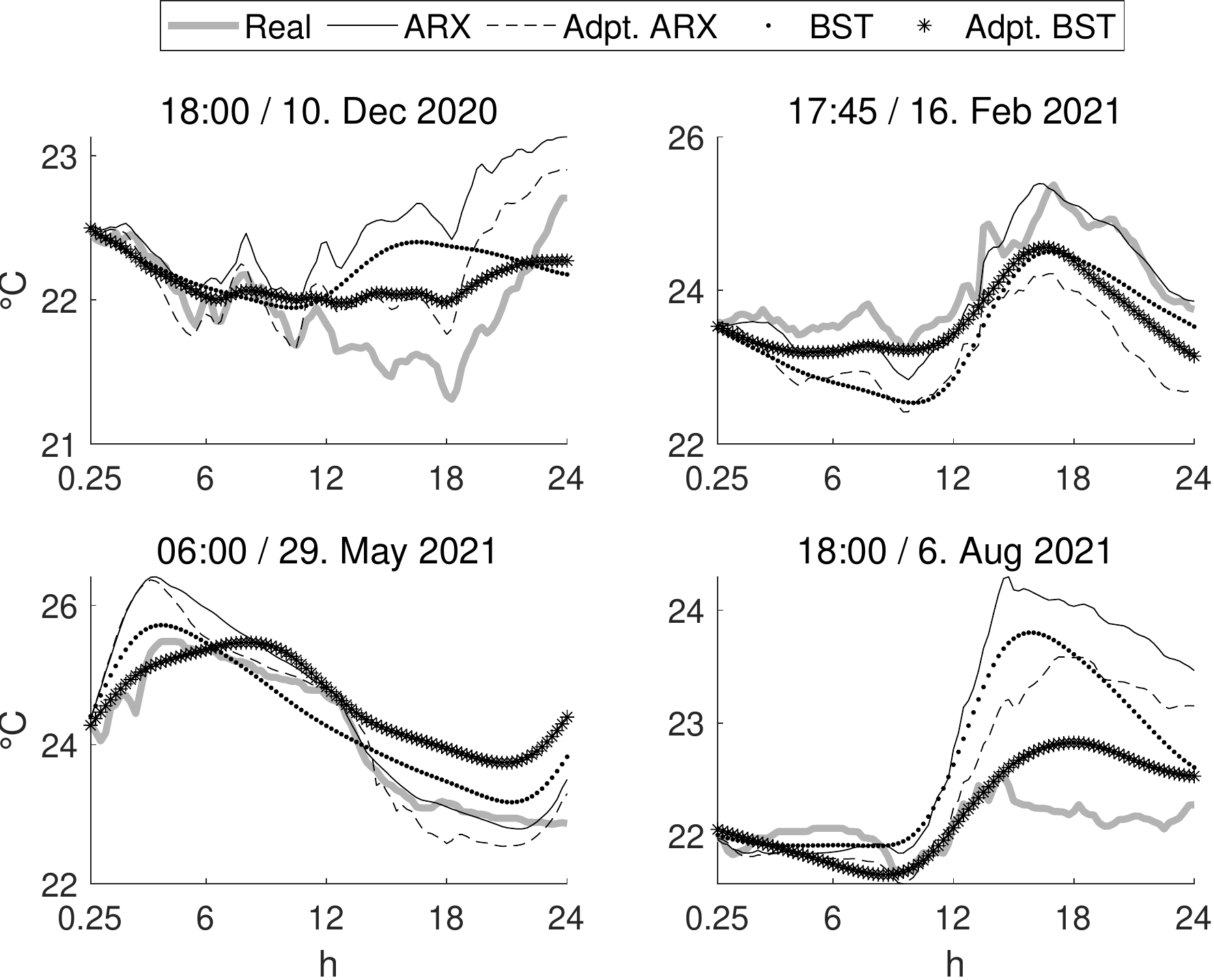}
    \caption{Selected zone temperature predictions at stated times over the horizon for Zürich building}
    \label{fig:ReaTraZurich}
\end{figure}

\begin{figure}[htp]
    \centering
    \includegraphics[width=12cm]{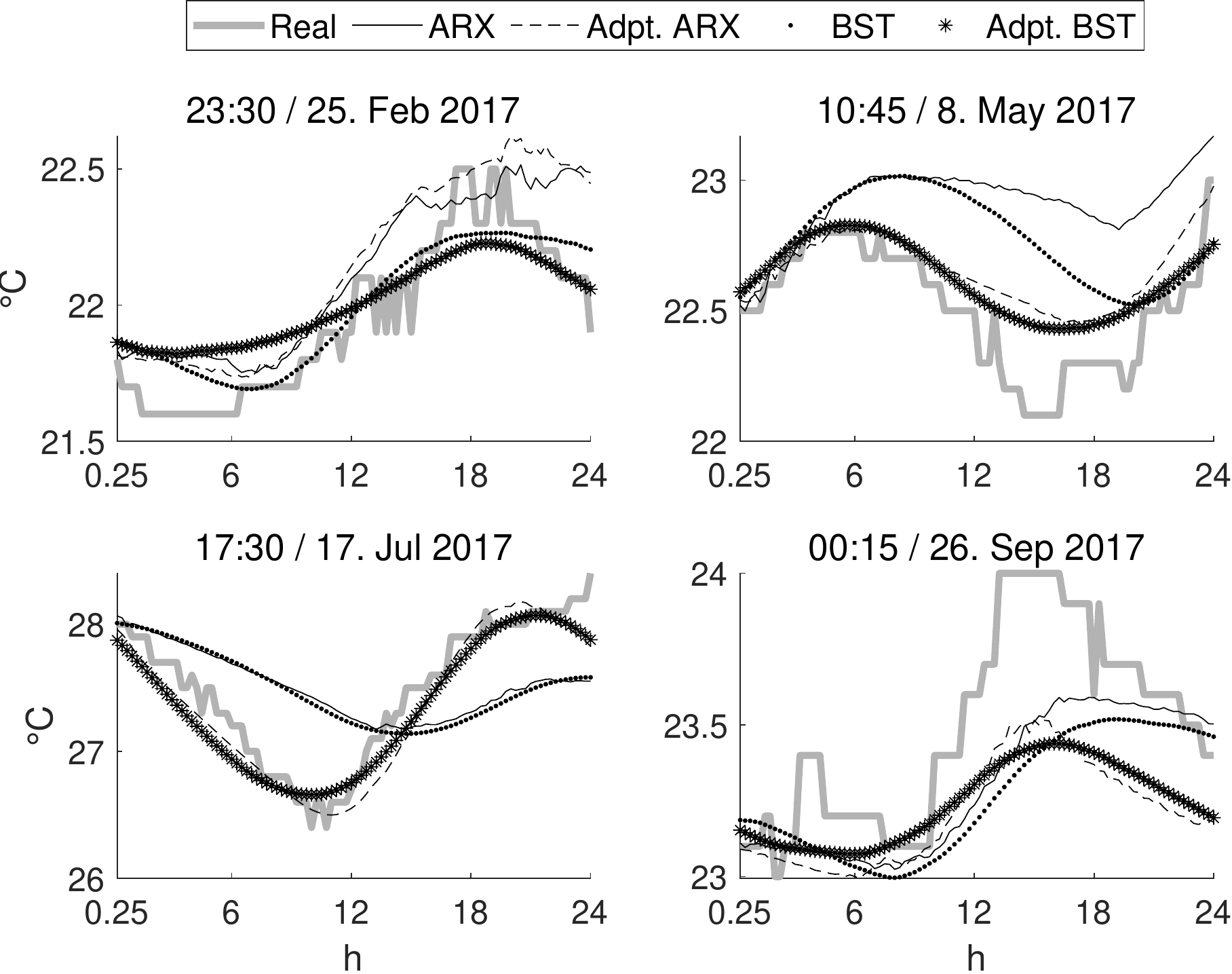}
    \caption{Selected zone temperature predictions at stated times over the horizon for Basel building}
    \label{fig:ReaTraBasel}
\end{figure}

Fig. \ref{fig:ReaTraZurich} and Fig. \ref{fig:ReaTraBasel} show a selection of specific zone temperature predictions for the Zürich and Basel buildings. We note the low temperature resolution in the latter. The following sections show a number of statistical analyses of all the data. Because of the different baselines, the two buildings are treated separately. 

\subsection{Zürich building}

\begin{figure}[htp]
    \centering
    \includegraphics[width=12cm]{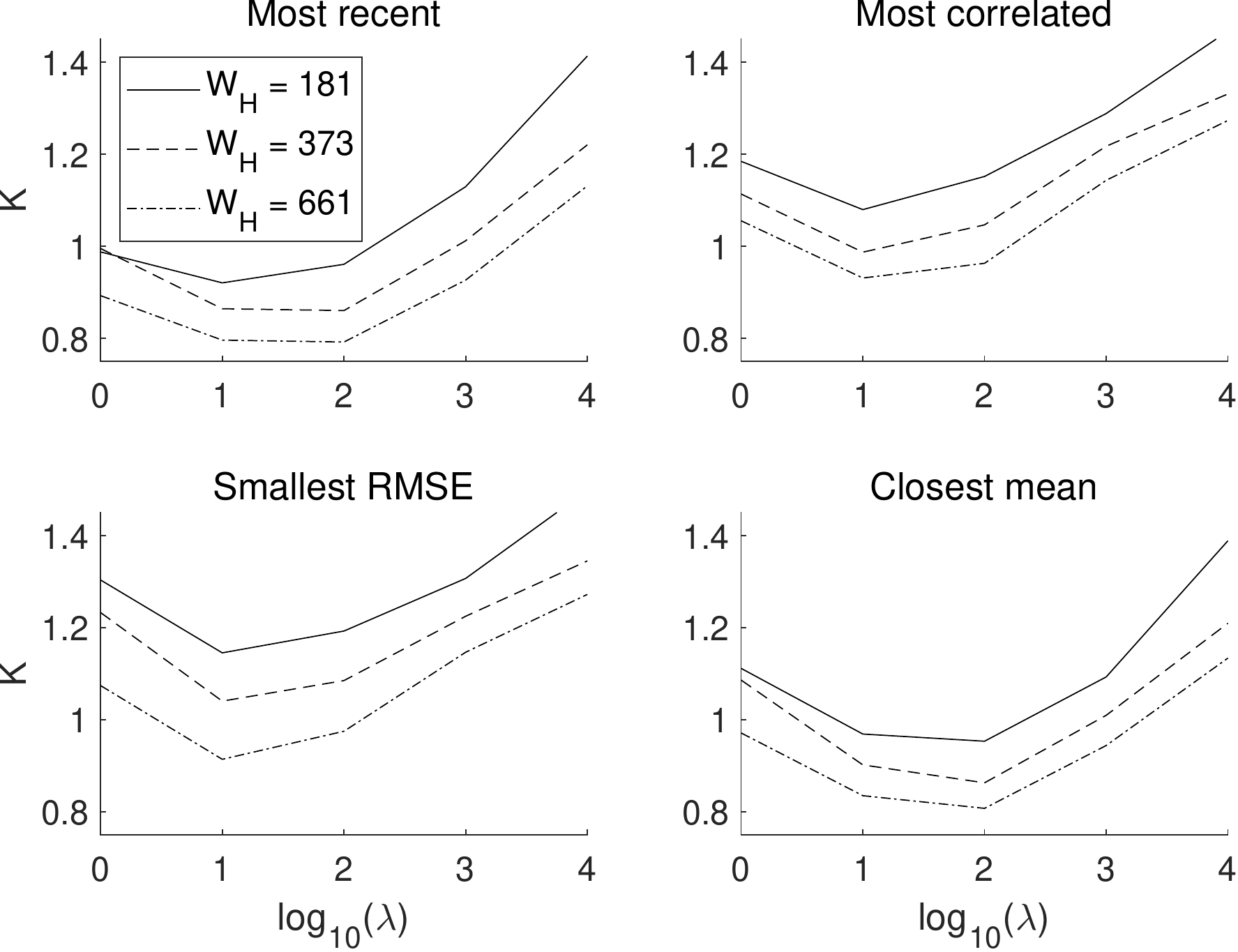}
    \caption{Comparison of mean temperature prediction error for adaptive BST, Zürich}
    \label{fig:ZurichAllDeePC}
\end{figure}

Fig. \ref{fig:ZurichAllDeePC} compares the mean prediction error during the evaluation phase for all variants of the adatpive BST. A larger matrix consistently leads to better performance. The \textit{Most recent} and \textit{Closest mean} variants perform better than the \textit{Most correlated} and \textit{Smallest RMSE} variants. 

\begin{figure}[htp]
    \centering
    \includegraphics[width=12cm]{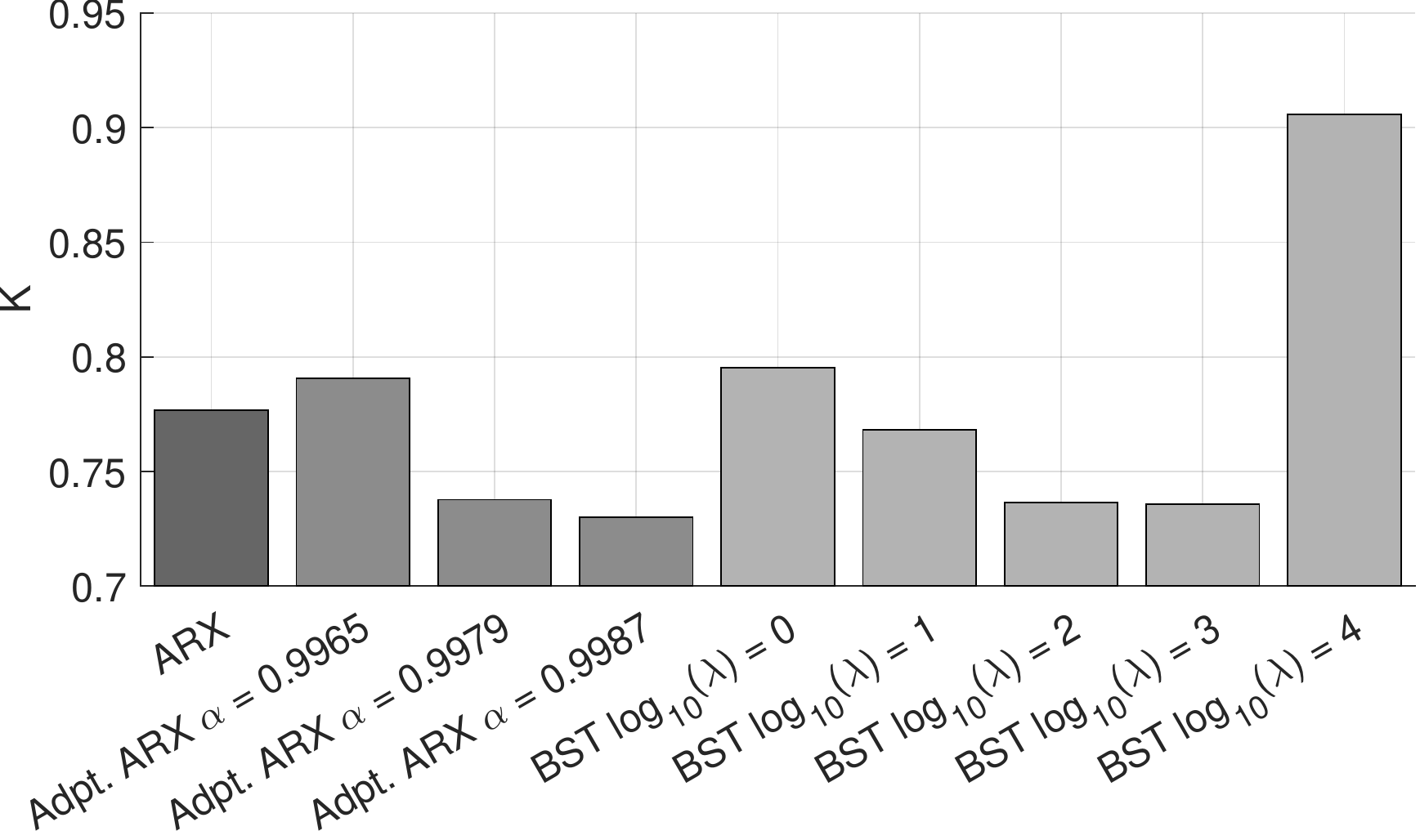}
    \caption{Comparison of mean temperature prediction error for reference controllers, Zürich}
    \label{fig:ZurichAllRef}
\end{figure}

Fig. \ref{fig:ZurichAllRef} shows all reference methods. The adaptive ARX with higher forgetting factors (i.e. a longer trace) perform best. The BST variants perform similarly to the ARX variants, depending on the regularization weight. 

For further comparison, the best variant of each of the four categories (ARX vs BST;  adaptive vs non-adaptive) is selected. For the adaptive BST, the best sub-variant is \textit{Most recent} with the widest trajectory matrix and the medium regularization weight. 

\begin{figure}[htp]
    \centering
    \includegraphics[width=8cm]{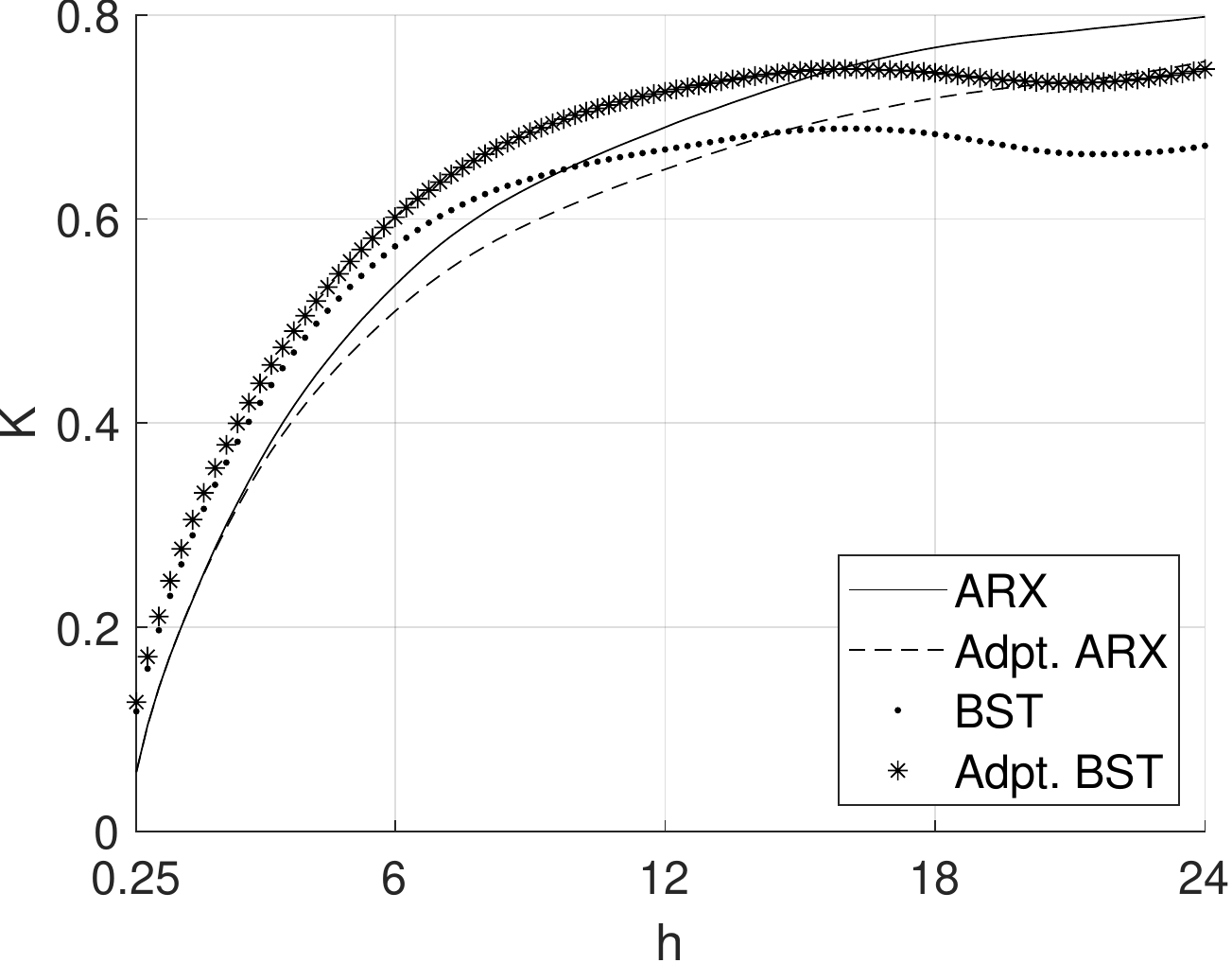}
    \caption{Mean prediction error over time, Zürich}
    \label{fig:ZurichTrajMean}
\end{figure}

\begin{figure}[htp]
    \centering
    \includegraphics[width=8cm]{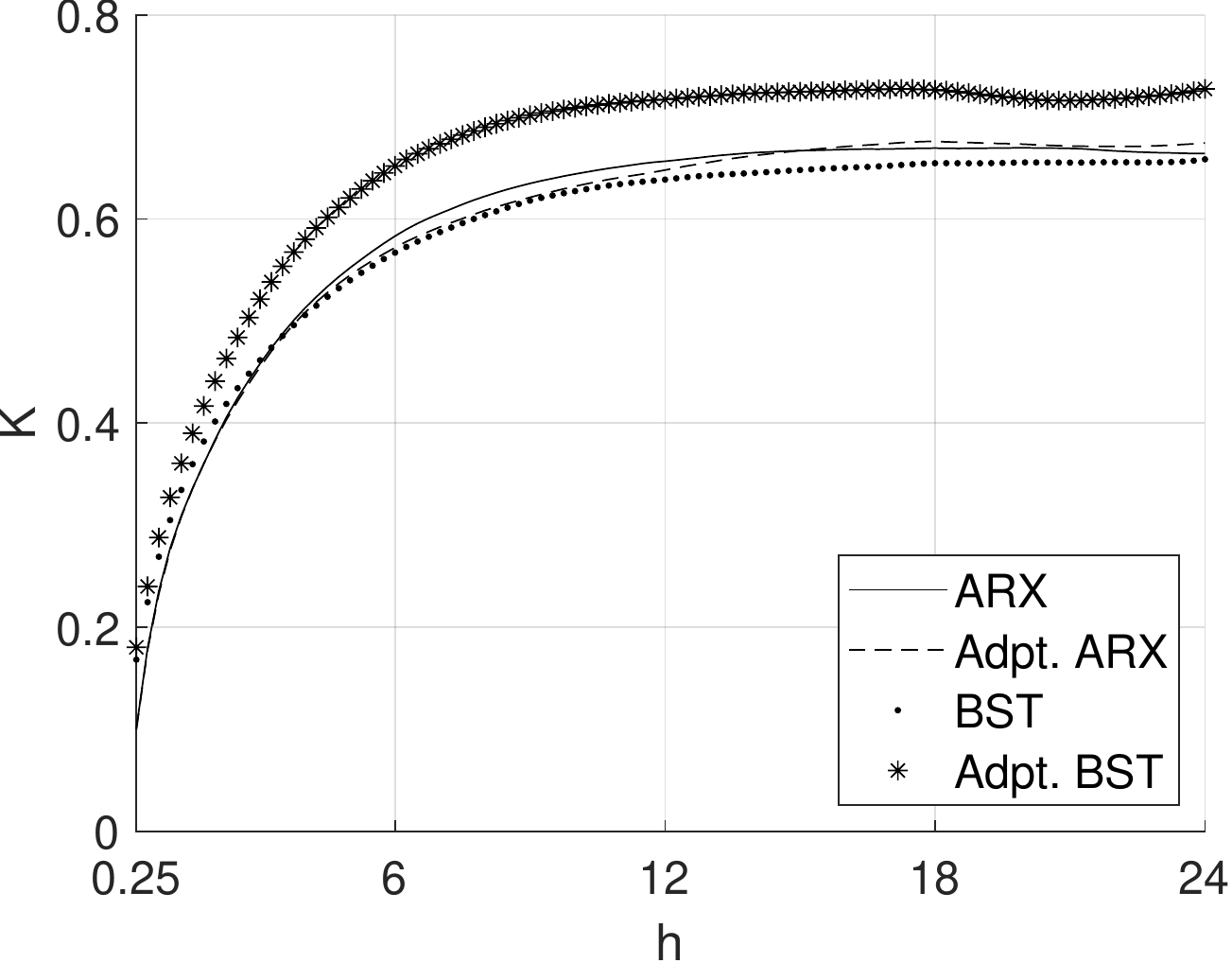}
    \caption{Standard deviation of prediction error over time, Zürich}
    \label{fig:ZurichTrajStd}
\end{figure}

Fig. \ref{fig:ZurichTrajMean} and Fig. \ref{fig:ZurichTrajStd} show the mean prediction errors over the horizon and the corresponding standard deviations. All variants perform similarly.  

\begin{figure}[htp]
    \centering
    \includegraphics[width=13cm]{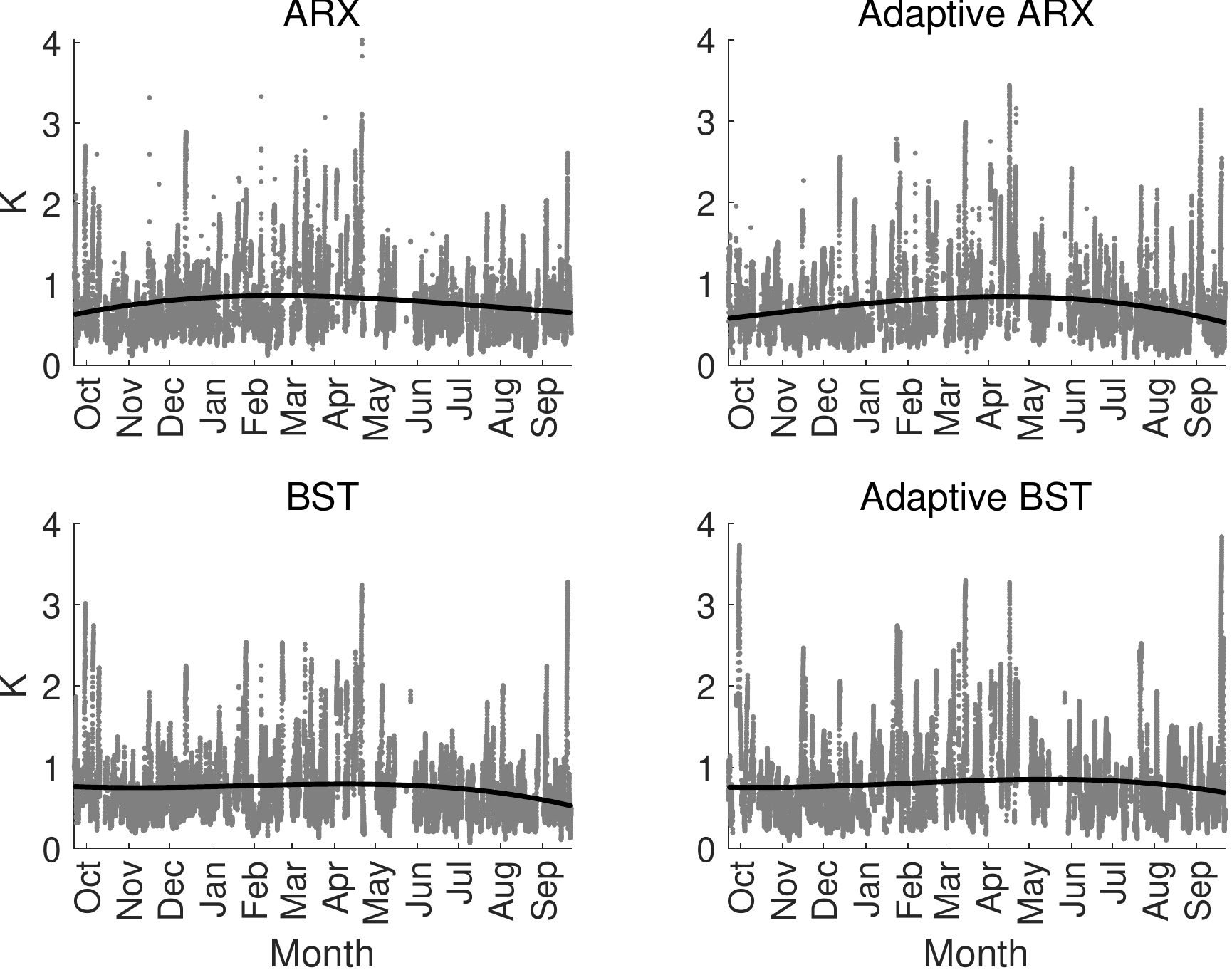}
    \caption{Seasonal prediction errors, Zürich}
    \label{fig:ErrAllZurich}
\end{figure}

\begin{figure}[htp]
    \centering
    \includegraphics[width=12cm]{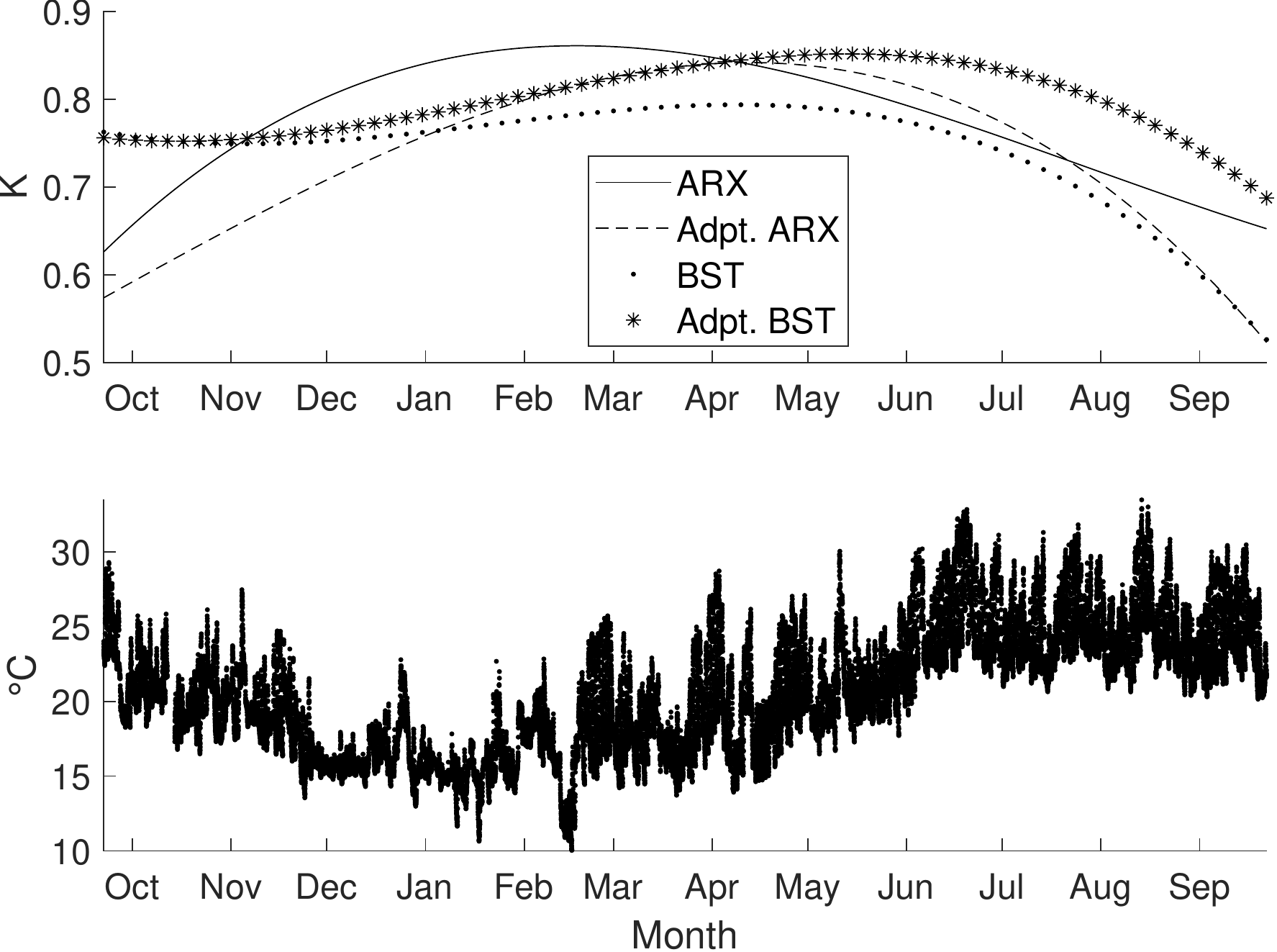}
    \caption{Seasonal prediction error fit comparison and ambient temperature, Zürich}
    \label{fig:ErrCompZurich}
\end{figure}

Fig. \ref{fig:ErrAllZurich} shows the prediction errors over the year and a cubic fit. Fig \ref{fig:ErrCompZurich} compares all fits to the ambient temperature. No clear pattern is recognizable. 

\subsection{Basel building}

\begin{figure}[htp]
    \centering
    \includegraphics[width=12cm]{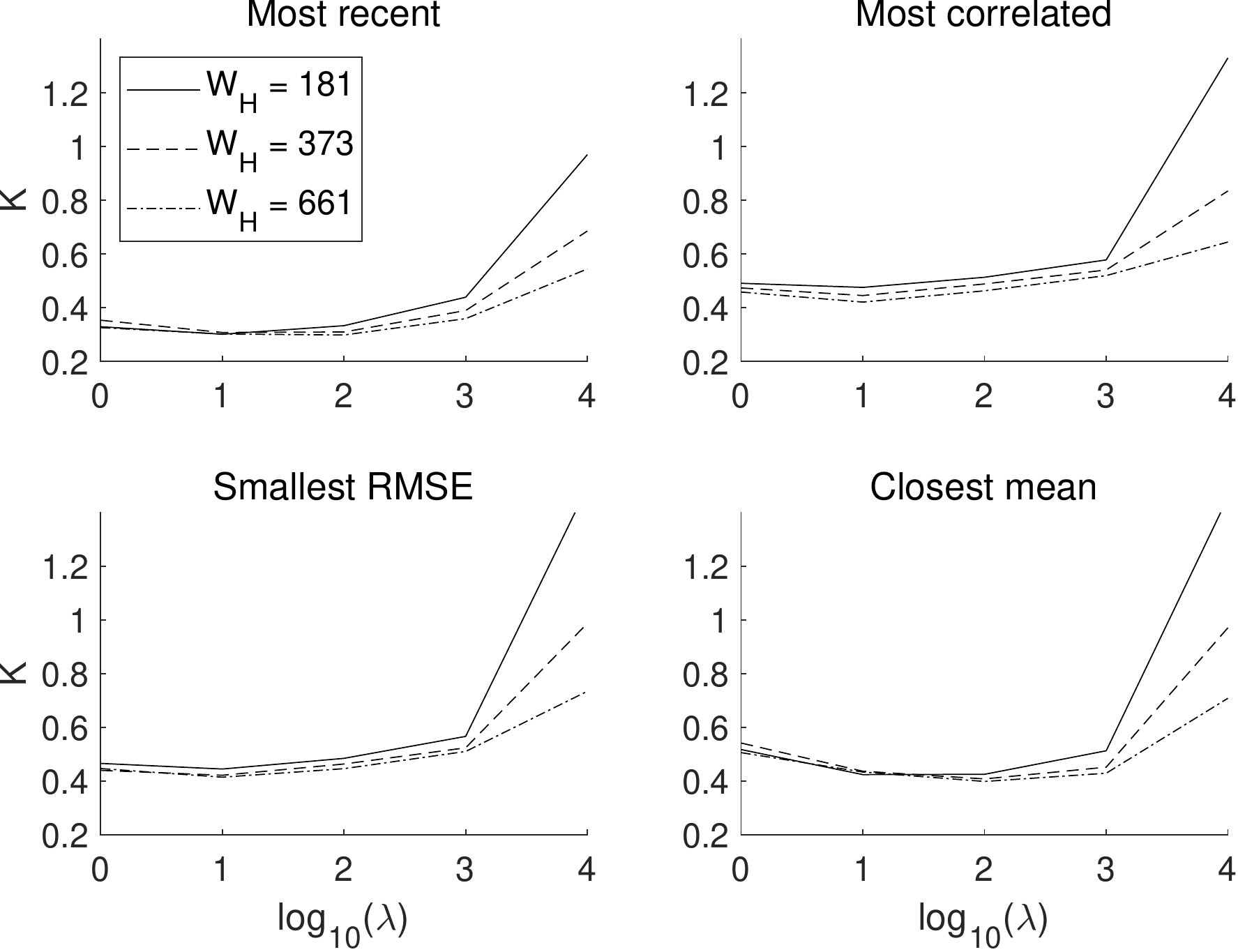}
    \caption{Comparison of mean temperature prediction error for adaptive BST, Basel}
    \label{fig:BaselAllDeePC}
\end{figure}

Fig. \ref{fig:BaselAllDeePC} compares the mean prediction error during the evaluation phase for all variants of the adatpive BST. As before, a wider trajectory matrix leads to better performance. The \textit{Most recent} variant with the medium regularization weight performs best. All other variants follow closely grouped. 

\begin{figure}[htp]
    \centering
    \includegraphics[width=12cm]{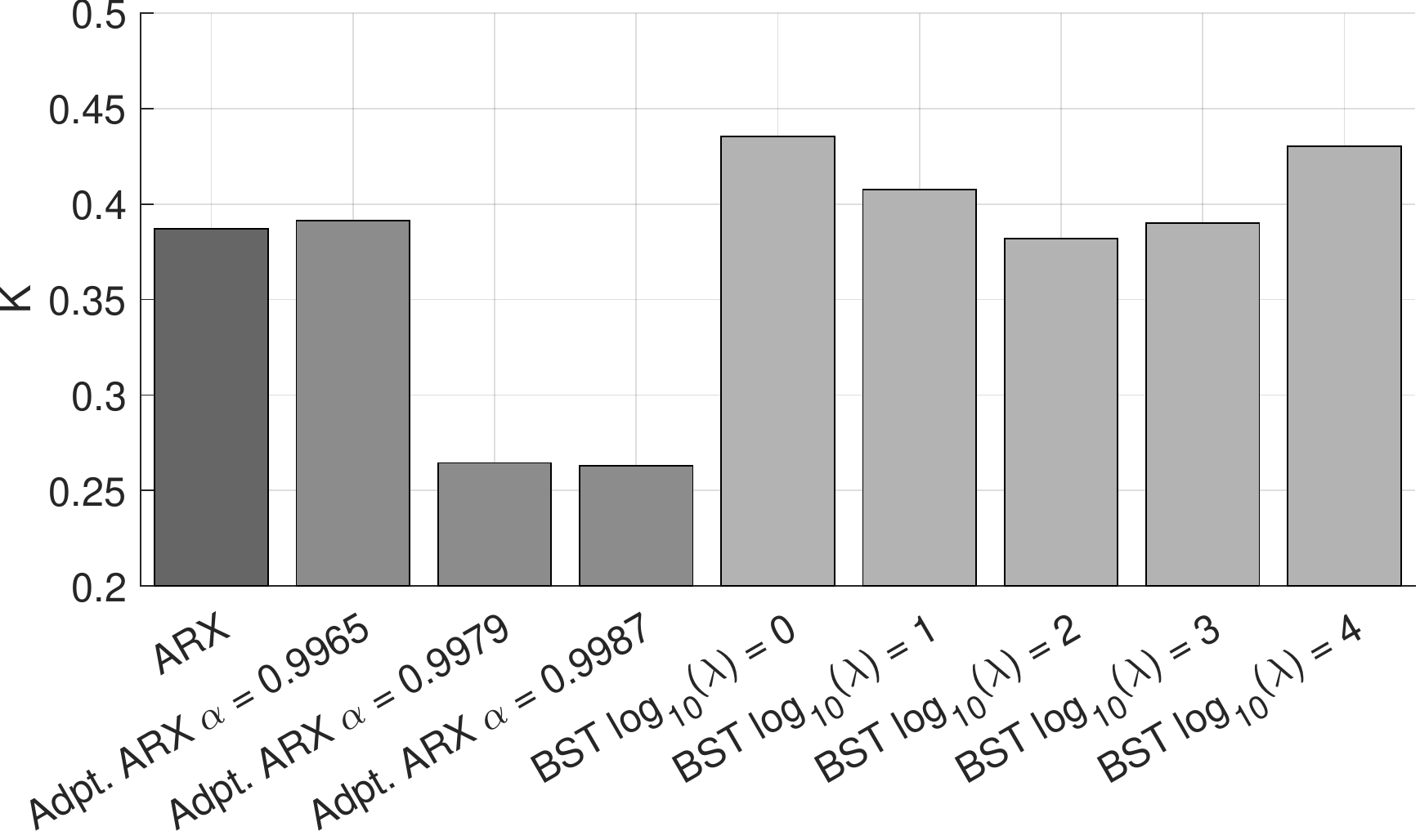}
    \caption{Comparison of mean temperature prediction error for reference controllers, Basel}
    \label{fig:BaselAllRef}
\end{figure}

Fig. \ref{fig:BaselAllRef} shows all reference methods. Two of the adaptive ARX variants perform significantly better than the non-adaptive one. The BST variants perform similarly to the ARX, depending on the regularization weight. 

As before, the best variant of each of the four principal categories is selected for comparison. 

\begin{figure}[htp]
    \centering
    \includegraphics[width=8cm]{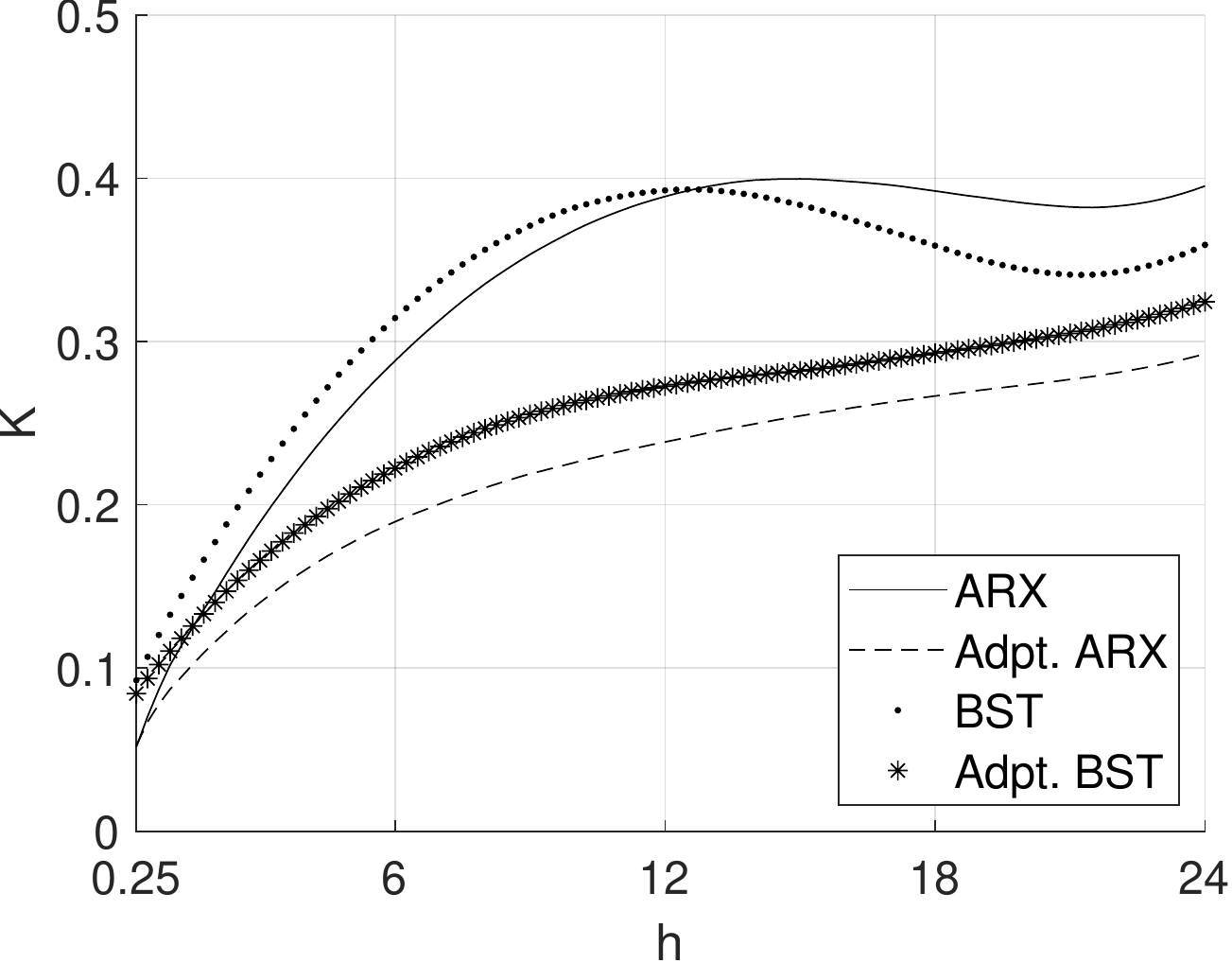}
    \caption{Mean prediction error over time, Basel}
    \label{fig:BaselTrajMean}
\end{figure}

\begin{figure}[htp]
    \centering
    \includegraphics[width=8cm]{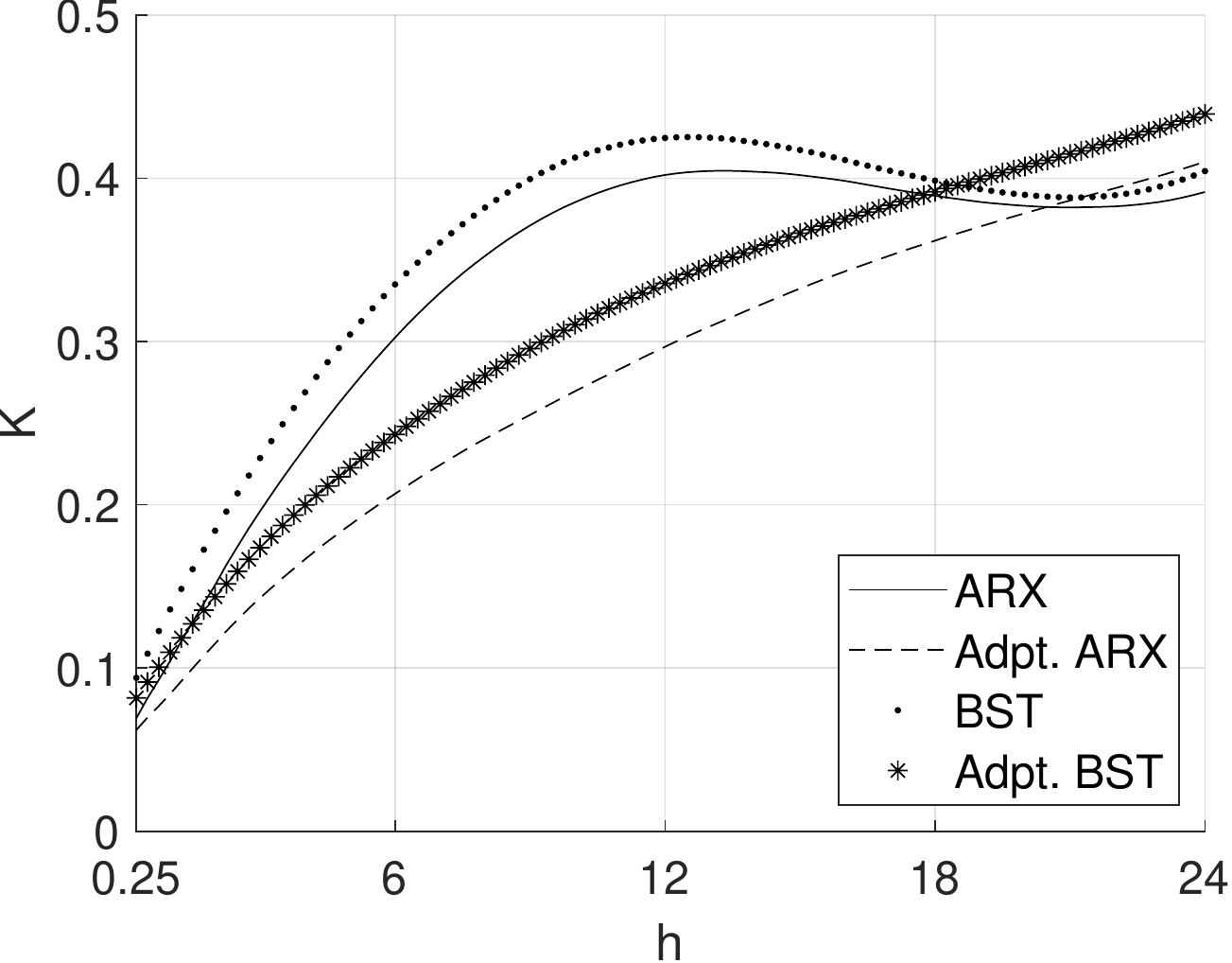}
    \caption{Standard deviation of prediction error over time, Basel}
    \label{fig:BaselTrajStd}
\end{figure}

Fig. \ref{fig:BaselTrajMean} and Fig. \ref{fig:BaselTrajStd} show the mean prediction errors over the horizon and the corresponding standard deviations. In general, the adaptive variants outperform the non-adaptive ones in this case, with a slight advantage for the ARX methods. 

\begin{figure}[htp]
    \centering
    \includegraphics[width=13cm]{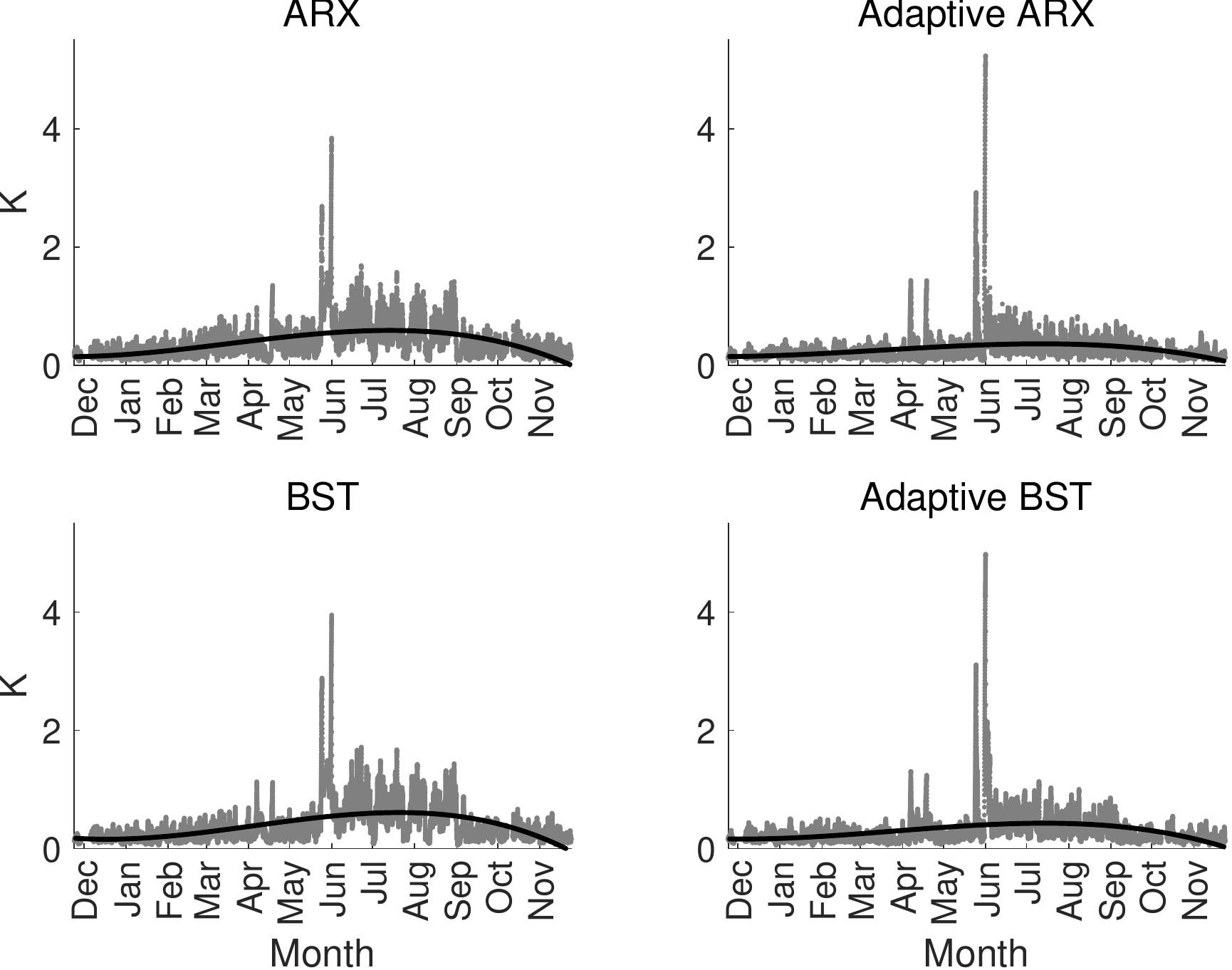}
    \caption{Seasonal prediction errors, Basel}
    \label{fig:ErrAllBasel}
\end{figure}

\begin{figure}[htp]
    \centering
    \includegraphics[width=12cm]{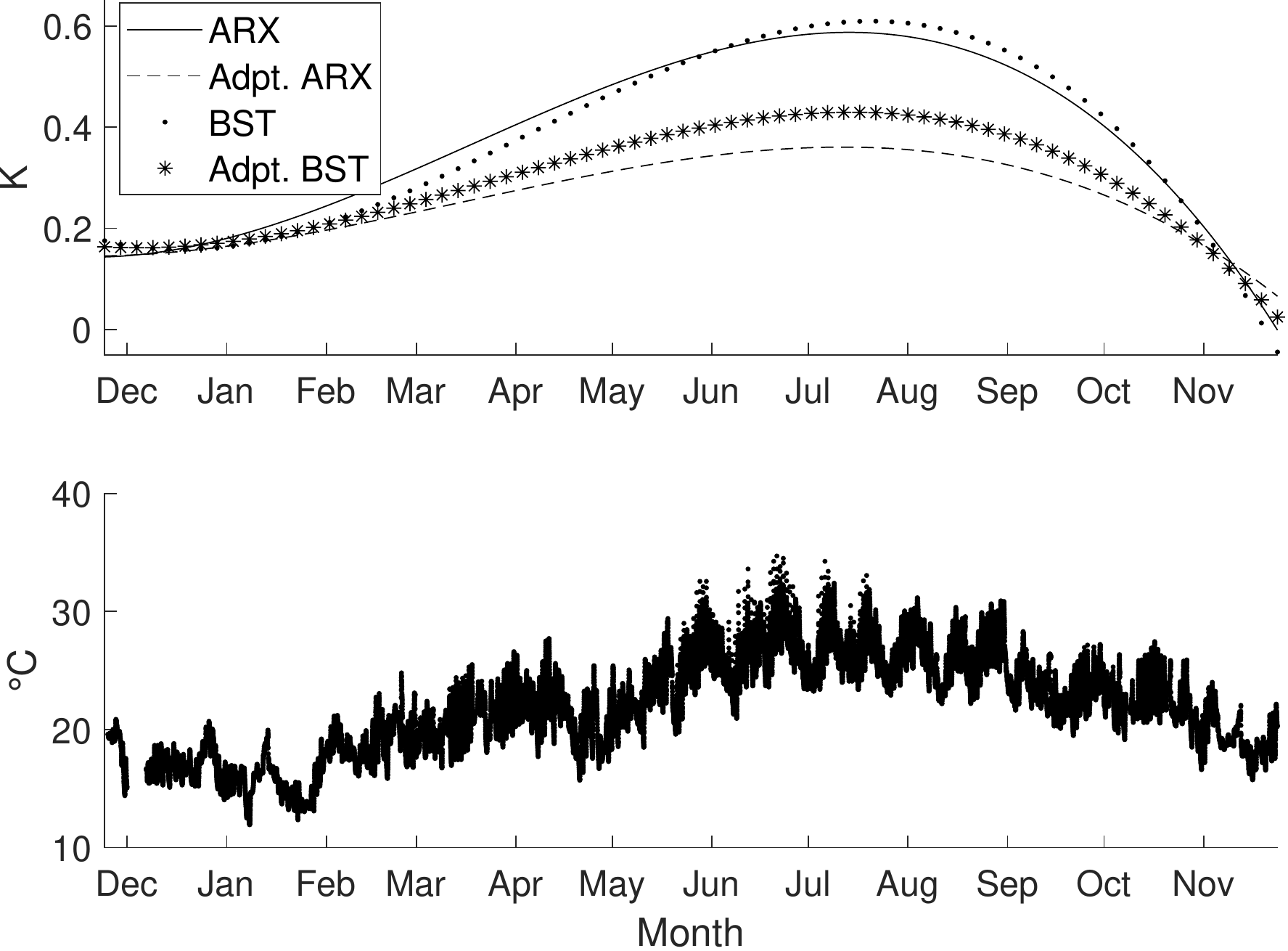}
    \caption{Seasonal prediction error fit comparison and ambient temperature, Basel}
    \label{fig:ErrCompBasel}
\end{figure}

Fig. \ref{fig:ErrAllBasel} shows the prediction errors over the year and a cubic fit. Fig. \ref{fig:ErrCompBasel} compares all fits to the ambient temperature. There are noticeable spikes in the errors around the middle of the simulation period for both adaptive and non-adaptive variants. The similarity between ARX and BST indicates an (unknown) physical cause, rather than an issue with the methods. The adaptive variants show significantly better performance during the summer months in Fig. \ref{fig:ErrCompBasel}. 

\subsection{A note on singular values of the trajectory matrices}

In dual control, the cost function often includes a term to maximize the level of excitation in the resulting trajectory \cite{klenskeApproximateDualControl2016, ebadatApplicationorientedInputDesign2017, hernandezvicenteStabilizingPredictiveControl2019, luRobustAdaptiveModel2020, zacekovaZoneMPCGuaranteed2020, bruggemannForwardlookingPersistentExcitation2022}. According to Willems' lemma, a loss of full row rank means a loss of persistent excitation in a Hankel matrix. The smallest singular value of a matrix can be thought of as a measure of how close it is to this. Therefore, we hypothesized a correlation between the smallest singular value of the trajectory matrix and the resulting prediction accuracy. However, no such correlation was found. 

\section{Discussion and Conclusions}

One limitation of the study is the small data set, consisting of two similar buildings in similar climate. A greater variety of building types and climates would have been desirable, but corresponding data sets were not available. 

The prediction error for the Basel building is noticeably smaller than for the Zurich building. This is likely due to the larger size and the smaller window-to-wall area fraction. Thus giving the Basel building a much higher inertia-to-disturbance ratio. 

The average prediction accuracy of the BST and ARX methods is shown to be similar, with a slight advantage for the ARX. However, the computation of ARX is faster by orders of magnitude. \cite{schwarzDataDrivenControlBuildings2020} and \cite{kerkhofOptimalControlAutonomous2020} report similar ratios for the computation time comparing DeePC to conventional MPC. Furthermore, ARX has fewer tuning parameters. Excluding the worst choices of the regularization weight $\lambda$ for BST, all methods are sufficiently accurate, from an application point of view, with prediction errors well below $1 K$ over a $24 h$ horizon. In more general terms, linear time-invariant methods are sufficient to predict the zone temperatures in the studied buildings. We note that this accuracy is achieved without intentional excitation of the system for the purpose of identification. 

ARX has a clear advantage in the early prediction steps, which fades toward the later prediction steps. Accuracy in the early prediction steps is more important for most applications of predictive control. However, a more consistent accuracy over a long horizon could be desirable for specific applications. For example, the day-ahead prediction of electricity consumption required for some forms of demand response \cite{qureshiModelPredictiveControl2014}. 

Comparing the different methods to select data for trajectory matrices, using the most recent data is equally or more accurate than selecting data based on similar weather. It is also the simplest method to implement and the fastest to compute. 

The adaptive and non-adaptive methods perform similarly, despite the data sets including seasonal transitions. However, one year is still a short span, relative to the time scales at which building materials significantly degrade \cite{AcceleratedAgeingDurability2016, ImpactBuildingEnvelope2017}. Furthermore, the data sets do not include any substantial changes in the occupancy, use or surroundings of the buildings. Accounting for such long-term changes does not require the continuous model updating methods used in this study. A periodic update as used in \cite{woliszSelflearningModelPredictive2020} should be sufficient. 

Regarding the gaps in the data, the simplest possible methods of simply omitting them for the BST case and skipping them for the ARX case are found to work well. 

\section{Declaration of competing interest}

The authors declare that they have no known competing financial interests or personal relationships that could have appeared to influence the work reported in this paper. 

\section{Data availability}

The data used in this study is confidential. 

\section{Acknowledgements}

We would like to thank the Interreg programme for their support. We would further like to thank Loris Di Natale from EMPA and Gregor Steinke from FHNW for providing the data sets used in this study, as well as Ralph Eismann from FHNW for his  advisory contributions.


\bibliography{mybibfile}

\end{document}